\documentclass[aip,
jcp,
amsmath,amssymb,
reprint,
dvipsnames
]{revtex4-1}

\usepackage{graphicx}
\usepackage{dcolumn}
\usepackage{bm}

\usepackage[utf8]{inputenc}
\usepackage[T1]{fontenc}
\usepackage{mathptmx}
\usepackage{import}

\usepackage{mwe}

\usepackage[normalem]{ulem}
\usepackage{xcolor}

\begin{document}

\title[]{Ground-state properties of the narrowest zigzag graphene nanoribbon from quantum Monte Carlo and comparison with density functional theory}

\author{Raghavendra Meena}
\email{raghavendra.meena@wur.nl}
\affiliation{Biobased Chemistry and Technology, Department of Agrotechnology and Food Sciences, Wageningen University and
Research, PO Box 17, 6700 AA Wageningen, The Netherlands.}
\affiliation{Laboratory of Organic Chemistry, Department of Agrotechnology and Food Sciences, Wageningen University and
Research, PO Box 17, 6700 AA Wageningen, The Netherlands.}
\author{Guanna Li}
\affiliation{Biobased Chemistry and Technology, Department of Agrotechnology and Food Sciences, Wageningen University and
Research, PO Box 17, 6700 AA Wageningen, The Netherlands.}
\affiliation{Laboratory of Organic Chemistry, Department of Agrotechnology and Food Sciences, Wageningen University and
Research, PO Box 17, 6700 AA Wageningen, The Netherlands.}
\author{Michele Casula}
\email{michele.casula@sorbonne-universite.fr}
\affiliation{Institut de Min\'eralogie, de Physique des Mat\'eriaux et de Cosmochimie (IMPMC), Sorbonne Université, CNRS UMR 7590, IRD UMR 206, MNHN, 4 Place Jussieu, 75252 Paris, France.}

\date{\today}

\begin{abstract}
By means of quantum Monte Carlo (QMC) calculations from first principles, we study the ground-state properties of the narrowest zigzag graphene nanoribbon, with an infinite linear acene structure. We show that this
quasi-one-dimensional system is 
correlated and its ground state is made of localized $\pi$ electrons whose spins are antiferromagnetically (AFM) ordered. The AFM stablization energy (36(3) meV per carbon atom) and the absolute 
magnetization (1.13(1) $\mu_\textrm{B}$ per unit cell)
predicted by QMC are sizable, and they suggest the survival of antiferromagnetic correlations above room temperature. These values can be reproduced to some extent by density functional theory (DFT) only by assuming 
strong interactions,
either within the DFT+U framework or using hybrid functionals. Based on our QMC results, we then provide the 
strength of Hubbard repulsion in DFT+U suitable for this class of systems.
\end{abstract}

\maketitle

\section{\label{intro}INTRODUCTION\protect\\}

Zigzag graphene nanoribbons (ZGNRs) have attracted a great deal of interest in the last decades as ones of the most promising graphene derivatives for technological applications, in particular for their potential use in spintronic devices. Indeed, density functional theory (DFT) calculations 
showed that these systems are spin gapped, with the occurrence of spin-polarized zigzag edges\cite{nakada_edge_1996} that become antiferromagnetically (AFM) ordered\cite{son_energy_2006,wu_electronic_2015,correa_braiding_2018,jiang_unique_2007}. It has been suggested that an in-plane electric field perpendicular to the graphene nanoribbon could lead to a half-metal\cite{son_half-metallic_2006}, where only one spin carries electric current. One can make use of such desirable properties in several technological applications such as ribbon-based spintronics, sensors, and storage devices\cite{wang_2021_graphene}, as long as magnetism in these materials is strong enough to survive at operating temperatures. 

\begin{figure}[ht]
\centering
\includegraphics[scale=0.2]{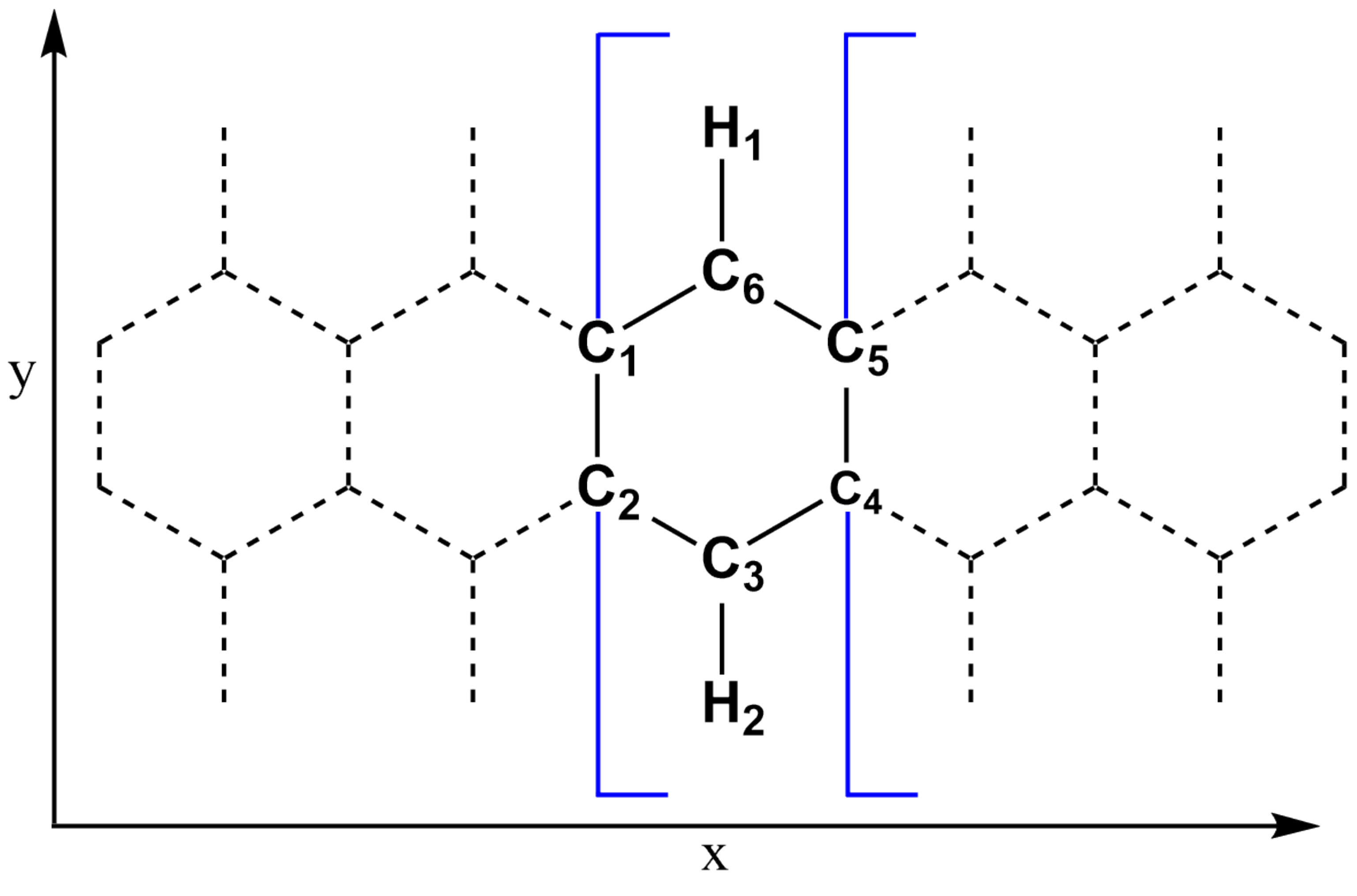}
\caption{\label{fig:unit-cell} Schematic view of the unit cell of 2-ZGNR, represented in square brackets.  This shows also the numbering of atoms, as used throughout the paper. The unit cell contains $N_\textrm{C}=4$ carbon atoms, and $N_\textrm{H}=2$ hydrogen atoms. The dashed lines are periodic replicas, extending to infinity on both sides.
}
\end{figure}

However, 
ZGNRs
are quasi-one-dimensional (quasi-1D) systems where 
electron-electron interaction
together with quantum fluctuations could play a major role in stabilizing different phases, competing each other in energy. This is particularly true for the narrowest nanoribbons, made of an infinite chain of fused benzene rings, such as to form a zigzag edge. The number of chains will determine the width of the ribbon. According to previous conventions\cite{fujita1996peculiar}, we will refer to the ribbon made of $n$ zigzag chains as an $n$-ZGNR. 
In the narrowest ribbon, i.e. the 2-ZGNR (see Fig.~\ref{fig:unit-cell}), the effects of quantum fluctuations are enhanced by low transverse dimensionality. A previous quantum Monte Carlo (QMC) study\cite{dupuy_fate_2018} on the polyacene molecules, the finite-length counterpart of 2-ZGNRs,
revealed how low dimensionality and
electron correlation favored a resonating-valence-bond (RVB) -like ground state, by penalizing solutions with spin polarized edges. Given these premises, a thorough verification of the DFT predictions needs to be carried out. Indeed, strong correlation and large quantum fluctuations, typical of 1D systems, are critical for DFT, which
could fail to predict the correct ground-state (GS) properties in these situations. 

A route to verify the DFT predictions
is
experimental. Unfortunately, ZGNRs 
are 
difficult to synthesize in a stable form because of their extremely reactive edges. Thus, a direct experimental demonstration of their 
GS
as an AFM-ordered phase is still to come, although significant progress has been made recently towards a controlled fabrication of these systems\cite{han2007energy,chen2007graphene,magda2014room,ruffieux_-surface_2016,fedotov2020excitonic,houtsma_atomically_2021,chen2021sub}. 
A gap opening in these samples has been experimentally detected by resistivity measurements\cite{han2007energy,chen2007graphene} and  scanning tunneling spectroscopy up to room temperature\cite{magda2014room}, and some degree of electron localization has been reported\cite{ruffieux_-surface_2016}. Still, the link to a possible AFM order could 
be made only
on the basis of DFT or model Hubbard Hamiltonians gap-size predictions.

Another possible way of verification is to perform first-principles calculations by using high-level theories, 
more accurate than 
standard
DFT but also computationally much more expensive.
Previous studies on the acene molecular series
revealed that the paramagnetic solution for their 
GS
is challenged by the competition against the formation of localized edge states, AFM coupled in an open-shell singlet configuration\cite{bendikov2004oligoacenes,hachmann2007radical,hajgato2009benchmark,chai2012density,rivero2013entanglement,yang2016nature,dupuy_fate_2018,schriber2018combined,trinquier2018qualitative}. Several correlated post-Hartree-Fock methods\cite{hachmann2007radical,hajgato2009benchmark,yang2016nature,schriber2018combined} and various types of DFT functionals\cite{bendikov2004oligoacenes,chai2012density} have been employed to 
unveil
the true 
GS, 
with results strongly dependent on the level of theory\cite{tonshoff2021pushing}. Moreover, in this class of systems, numerical DFT predictions quantitatively depend on the chosen functional.
On the other hand, QMC in its diffusion Monte Carlo implementation is one of the most accurate \emph{ab initio} theories, and it can be utilized to provide benchmark results in such 
correlated systems\cite{foulkes_quantum_2001,Wagner_2016,motta2020ground}.
This is the way we will purse in this paper in order to study the
GS
properties of the 2-ZGNR. 

Various other attempts have been made to model strong correlations in graphene and graphene derivatives using Hubbard Hamiltonians\cite{raghu_structural_2002,srinivasan_structural_1998} on a honeycomb lattice. However, a relevant question still debated
in this framework is about the correct value of Hubbard U repulsion one has to use to reproduce the physics taking place in a more realistic setting\cite{jung2011nonlocal,wehling2011strength,schuler2013optimal,csacsiouglu2017strength,PhysRevB.98.205123}. 
At variance with model Hamiltonian approaches\cite{golor2013quantum}, which suffer from this issue, our QMC framework
includes electron correlations entirely from first principles. By carrying out extensive variational and diffusion QMC calculations, we show that the AFM phase is the prime candidate for the $n$-ZGNR
GS, for $n=2$ and presumably for larger $n$.
We then compare the results obtained at the QMC level with 
the 
widely used DFT
to validate the use of certain DFT functionals, such as the hybrid PBE0, HSE and GauPBE 
schemes,
and determine the 
optimal
value of U to be used in the DFT+U framework.

The paper is organized as follows. Secs.~\ref{QMC-methods} and \ref{DFT-methods} introduce the QMC and DFT methods used in this work, respectively. Sec.~\ref{QMC-results} reports the ground-state properties obtained by QMC, while in Sec.~\ref{DFT-results} various DFT functionals are compared against QMC in order to find the ones that best reproduce the physics of 2-ZGNRs. Finally, Sec.~\ref{conclusions} draws the conclusions.

\section{METHODS}
\label{methods}

\subsection{QMC}
\label{QMC-methods}

All the QMC calculations were done using the TurboRVB package\cite{nakano_turborvb_2020}, a complete suite of \emph{ab initio} codes based on real-space Jastrow-correlated wave functions $\Psi$. 
With the aim of unveiling the ground-state (GS) nature of the narrowest zigzag graphene nanoribbon, we compared the variational Monte Carlo (VMC) energies ($E_\textrm{VMC}$) of two types of wave functions:
\begin{enumerate}
\item A Jastrow-correlated single-determinant wave function, describing a paramagnetic (PM) state;
\item A wave function with spin-broken symmetry in both the Jastrow and the determinantal part with $z$-projected total spin $\hat{S}_z=0$, developing an AFM long-range order.
\end{enumerate}
To reach the closest solution to the true GS, by means of the lattice regularized diffusion Monte Carlo (LRDMC)\cite{casula_diffusion_2005} method
we 
projected these wave functions to the lowest energy state $\Psi_\textrm{FN}$ compatible with the fixed node (FN) approximation\cite{reynolds_fixednode_1982}. The closer the nodes of the variational wave function to the GS ones, the lower the corresponding FN energy is. Thus, by comparing the FN energies ($E_\textrm{FN}$) of the two wave functions, one can determine what is the symmetry of the state that best represents the GS.

The two different PM and AFM many-body wave functions used in this work can be written in a unified form as:
\begin{equation}
    \Psi^\textbf{k}(\textbf{R})=\exp\{-U(\textbf{R},\Sigma)\} \det\{\phi^{\textbf{k},\uparrow}_j(\textbf{r}^\uparrow_i)\}\det\{\phi^{\textbf{k},\downarrow}_j(\textbf{r}^\downarrow_i)\},
    \label{wf}
\end{equation}
for $i,j \in \{1,\ldots,N/2\}$, where $N$ is the number of electrons in the simulation supercell, $\textbf{k}$ is the twist belonging to the supercell Brillouin zone,
$U$ is the Jastrow function, $\phi^{\textbf{k},\sigma}_i$ are one-body orbitals, 
 $\textbf{R}=\{\textbf{r}^\uparrow_1,\ldots,\textbf{r}^\uparrow_{N/2},\textbf{r}^\downarrow_1,\ldots,\textbf{r}^\downarrow_{N/2}\}$ is the $N$-electron spacial coordinate, and $\Sigma$ is the $N$-electron spin configuration. For the paramagnetic case, both Jastrow function and one-body orbitals do not depend on $\Sigma$, i.e. $U(\textbf{R},\Sigma)=U(\textbf{R})$ and $\phi^{\textbf{k},\uparrow}_i=\phi^{\textbf{k},\downarrow}_i=\phi^\textbf{k}_i$. Otherwise, a spin-broken solution can be stabilized. 

The Jastrow function $U$ is split into electron-nucleus, electron-electron, and electron-electron-nucleus parts:
\begin{equation}
U=U_{en}+U_{ee}+U_{een}.
\label{JasU}
\end{equation}
The electron-nucleus function has an exponential form and it reads as 
\begin{equation}
U_{en}=\sum_{iI} J_{1b}(r_{iI}), 
\label{JasUen}
\end{equation}
where the index $i$($I$) runs over electrons (nucleus), $r_{iI}$ is the electron-nucleus distance, and
\begin{equation}
J_{1b}(r)=\alpha \left( 1-e^{-r/\alpha} \right),
\label{J1b}
\end{equation}
with $\alpha$ a variational parameter. $J_{1b}$ cures the nuclear cusp conditions, and allows the use of the bare Coulomb potential for the hydrogen atom in our QMC framework. For the carbon atom, we used instead the Burkatzki-Filippi-Dolg (BFD) pseudopotential\cite{burkatzki_energy-consistent_2007}, to replace the $1s$ core electrons. The electron-electron function has a Pad\'e form and it reads as 
\begin{equation}
U_{ee}=-\sum_{ij} J_{2b}(r_{ij},\sigma_i,\sigma_j),
\label{JasUee}
\end{equation}
where the indices $i$ and $j$ run over electrons, $r_{ij}$ is the electron-electron distance, $\sigma_i$ is the spin of the $i$-th electron, and
\begin{eqnarray}
J_{2b}(r,\sigma,\bar{\sigma}) & = & 0.5~ r/(1+ \beta r), 
\label{J2b_anti}
\\ J_{2b}(r,\sigma,\sigma) & = & 0.25 ~r/(1+ \beta r), 
\label{J2b_para}
\end{eqnarray}
with $\beta$ a variational parameter. This two-body Jastrow term fulfills the cusp conditions for both parallel $(\sigma_i=\sigma_j=\sigma)$ and antiparallel $(\sigma_i=\sigma, \sigma_i=\bar{\sigma})$ spins. The last term in the Jastrow factor is the electron-electron-nucleus function: 
\begin{equation}
U_{een}=\sum_{ijI} \sum_{\gamma\delta} M_{\gamma \delta I}(\sigma_i,\sigma_j)\chi_{\gamma I}(r_{iI}) \chi_{\delta I}(r_{jI}),
\label{JasUeen}
\end{equation}
with $M_{\gamma \delta I}(\sigma,\sigma^\prime)$ a matrix of variational parameters symmetric in both orbital $(\gamma,\delta)$ and spin $(\sigma,\sigma^\prime)$ sectors, and $\chi_{\gamma I}(r)$ a $(2s,2p,1d)$ Gaussian basis set, with orbital index $\gamma$, centered on the nucleus $I$. Note that the sum in Eq.~\ref{JasUeen} runs over a single nuclear index $I$. Thus, in the variational wave function we neglect inhomogeneous correlations between electrons belonging to different sites, in order to reduce the number of relevant parameters. The $J_{1b}$ and $J_{2b}$ Jastrow functions have been periodized using a $\textbf{r} \rightarrow \textbf{r}^\prime$ mapping that makes the distances diverge at the border of the supercell, as explained in Ref~\onlinecite{nakano_turborvb_2020}. For the inhomogeneous $U_{een}$ part, the Gaussian basis set $\chi$ has been made periodic by summing over replicas translated by lattice vectors. The Gaussian basis set of the Jastrow factor is constructed such that the radial part of the second-$s$ and second-$p$ Gaussian functions have a $r^2 e^{-Z r^2}$ form, to enforce orthogonality in the basis set. For the paramagnetic system, $M_{\gamma \delta I}(\sigma_i,\sigma_j)$ is taken spin independent, with the same entries for all spin sectors. Similarly, $J_{2b}(r,\sigma,\sigma^\prime)=0.5 r/(1+ \beta r) ~~ \forall \sigma, \forall \sigma^\prime$, namely $J_{2b}$ will fulfill only the electron cusps conditions for antiparallel spins in the spin-independent case.

The one-body orbitals $\phi$ are expanded onto a primitive $(8s,8p,2d,1f)$ Gaussian basis set for carbon (C), $(4s,2p,1d)$ for hydrogen (H), which we contracted into an optimal set of 9 hybrid orbitals for C and 2 hybrid orbitals for H (9C\&2H), by using the geminal embedding orbitals (GEO) contraction scheme\cite{sorella_geminal_2015} at the $\Gamma$ point. $\phi$s' are made periodic by using the same scheme as for the $\chi$s'.

The best orbital contractions are determined at the DFT level within the local density approximation (LDA), for the system in the paramagnetic phase.
The TurboRVB package allows one to perform LDA-DFT calculations in a Gaussian basis set, and with a mean-field wave function as in Eq.~\ref{wf} supplemented with the one-body Jastrow factor only, namely with $U=U_{en}$ in Eq.~\ref{JasU}.
Note that thanks to the one-body Jastrow factor included in the DFT wave function, it is possible to use the \emph{ab initio} Hamiltonian with bare Coulomb potential for the H atom. Not only the fulfillment of electron-ion cusp conditions\cite{pack_cusp_1966} via the Jastrow factor allows the use of divergent electron-ion potentials in the Hamiltonian, but it also significantly speeds up the basis set convergence already at the DFT level. Thus, it avoids the use of large Gaussian exponents, which are then cut off from the regular basis set. By employing a primitive Gaussian basis set of cc-pVTZ\cite{kendall_electron_1992,davidson_comment_1996} quality, optimized for the C pseudopotential\cite{burkatzki_energy-consistent_2007}, we determined the optimal GEO-contracted orbitals by requiring a convergence within 0.5 mH from the cc-pVTZ energy. It turns out that the 9C\&2H GEO contractions (see Tab.~\ref{tab:geo-contractions}) yield the desired accuracy at a less computational expense and with less variational parameters in the subsequent QMC calculations.

\begin{table}[ht]
\caption{\label{tab:geo-contractions} 
2-ZGNR energies from LDA-DFT $\Gamma$-point calculations for the primitive cc-pVTZ and GEO contracted basis sets. Sub-mHa convergence is reached with a 9C\&2H GEO contraction.
}
\begin{tabular}{ccc}
\hline
\hline
Basis set & Energy per C (Ha) & Overlap with cc-pVTZ \\
\hline
\hline
cc-pVTZ   & -5.95858               & 1.00000              \\
4C\&2H    & -5.75062               & 0.89992              \\
7C\&2H    & -5.95597               & 0.99927              \\
9C\&2H    & -5.95802               & 0.99983              \\
\hline
\hline
\end{tabular}
\end{table}

Beside the basis set error, which is usually very mild in LRDMC calculations, another source of bias in QMC 
energies
of extended systems comes from finite-size (FS) errors. In this work, the 2-ZGNR is put in a 3-dimensional periodic box, aligned along the $x$-axis and with vacuum separation of 7.2 \AA\ and 7 \AA\ in $y$ and $z$ directions, respectively, to avoid interaction between periodic images (see App.~\ref{DFT_appendix} and Fig.~\ref{fig:vac-z}). We used twisted boundary conditions for box-periodicity\cite{lin2001twist}.
Finite-size scaling analysis has been done by extending the unit cell in the $x$ direction (see Fig.~\ref{fig:unit-cell}).
We controlled FS errors by using the exact special twist method\cite{dagrada_exact_2016}. The special twist is the $\textbf{k}_s$-point in the first Brillouin zone (IBZ) where the energy is equal (or numerically very close) to the energy averaged over the full IBZ. Thus, it is very similar in spirit to the Baldereschi $\textbf{k}$-point\cite{baldereschi_mean-value_1973}, which is instead determined by symmetry arguments only. This procedure is ``exact'' in DFT or in other one-body theories, while it is an approximation in QMC, since the latter is affected by both one-body and two-body FS errors\cite{kent_finite-size_1999,kwee_finite-size_2008}. However, it has been proven that QMC calculations performed at the special twist show a smooth energy dependence on the supercell size\cite{dagrada_exact_2016}.

We carried out LDA-DFT calculations in the Gaussian GEO-contracted basis set to determine the special twist at a given supercell size in the PM phase. The quasi-1D nature of the 2-ZGNRs restricts this search to $\textbf{k}_s=(k_x,0,0)$, for a ribbon oriented along the $x$ direction. The $k_x$ found for different supercell sizes is reported in Tab.~\ref{tab:pm-QMCenergies}. We used the same $\textbf{k}_s$ in the AFM phase (Tab.~\ref{tab:afm-QMCenergies}).

Final LDA-DFT calculations at the special $\textbf{k}_s$-point are performed to find the self-consistent Kohn-Sham orbitals, plugged in the variational wave function of Eq.~\ref{wf} as starting $\phi^{\textbf{k}_s,\sigma}_i$ one-body orbitals. These calculations are performed at both LDA and local spin density approximation (LSDA) levels. The LDA-DFT does not 
break the spin symmetry and initializes the one-body orbitals of the PM wave function. For the LSDA-DFT, we impose an initial magnetic field which stabilizes an AFM pattern, and switched off after the first iteration, in order to initialize the orbitals of the AFM wave function.

To correlate the Slater determinant generated by DFT, we applied the Jastrow factor leading to the QMC wave function of Eq.~\ref{wf}. In order to obtain the best trial wave function in the variational sense (i.e. as close as possible to the true ground-state), we optimize\cite{sorella_wave_2005,umrigar2007alleviation} the variational parameters by minimizing the QMC energy of the variational wave function $\Psi$. At the first QMC step, the $\alpha$, $\beta$ and $M_{\gamma \delta I}(\sigma,\sigma^\prime)$ linear coefficients of the Jastrow part are optimized. Then, the Gaussian exponents in the Jastrow basis set $\chi$ are relaxed as well. Afterwards, the linear coefficients in $\phi^{\textbf{k}_s,\sigma}_i$ and the Gaussian exponents of the determinantal part are also optimized, simultaneously with the Jastrow factor. As a last step, the 2-ZGNR geometry is relaxed together with the wave function, thanks to the latest progress in the QMC calculations of ionic forces\cite{doi:10.1063/1.3516208}.

After fully optimizing the QMC trial wavefunctions, we performed 
VMC simulations to evaluate the variational energy $E_\textrm{VMC}$ for both paramagnetic and AFM states. Finally, we carried out
LRDMC calculations based on the optimized variational wave functions. The LRDMC is a lattice regularized version of the standard diffusion Monte Carlo (DMC) algorithm\cite{casula_diffusion_2005}, obtained by discretizing the Laplacian on an adaptive mesh\cite{nakano2020speeding}. We used a discretization step of 0.25 Bohr, which guarantees unbiased energy differences. Like DMC\cite{reynolds_fixednode_1982}, the LRDMC is a stochastic projection method to solve the imaginary-time Schr\"odinger equation, within the constraints imposed by the FN approximation. Moreover, LRDMC is fully compatible with the use of non-local pseudopotentials, such as the one for the carbon atom. 
The wave function projection is done through a multi-walker branching algorithm with fixed number $n_w$ of walkers\cite{buonaura1998numerical}. We used $n_w=9216$ for the largest system, with 50 hops per walker between two branching steps, and less than 0.5\% of the walkers population killed per step. The residual finite population bias is then corrected by the correcting factors\cite{buonaura1998numerical}.
At convergence, the LRDMC algorithm samples the mixed distribution $\Psi_\textrm{FN}\Psi$, being $\Psi_\textrm{FN}$ the projected wave function. Thus, one evaluates the FN energy as $E_\textrm{FN}=\langle \Psi_\textrm{FN} | H_\textrm{FN} | \Psi \rangle / \langle \Psi_\textrm{FN} | \Psi \rangle $. $H_\textrm{FN}$ is the effective FN Hamiltonian implementing the FN constraints, 
whose GS is exactly $\Psi_\textrm{FN}$.
The variational character of $E_\textrm{FN}$ implies that $E_0 \le E_\textrm{FN} < E_\textrm{VMC} $, where $E_0$ is the unknown true GS energy of the system, and $E_\textrm{VMC}$ is the energy of the wave function $\Psi$ before projection. The difference between $E_0$ and $E_\textrm{FN}$ is the FN error.

The above steps (wave function initialization, optimization and projection) are repeated for different supercell sizes, in order to reduce many-body FS errors. For each phase, we employed 3 supercells, made of 12 C (3-ring periodic), 24 C (6-ring periodic) and 36 C (9-ring periodic), respectively. The VMC and LRDMC energies of these 3 supercells are then corrected by applying the Kwee-Zhang-Krakauer (KZK)\cite{kwee_finite-size_2008} energy functional. The KZK-corrected energies, reported in Tabs.~\ref{tab:pm-QMCenergies} and \ref{tab:afm-QMCenergies}, are finally extrapolated to the thermodynamic limit by performing a linear fit in $1/N$ (see Fig.~\ref{fig:FS-total-energy}). The extrapolated energies are thus FS error free. Not only the energies but also other physical observables, such as the absolute magnetization, can be extrapolated to their thermodynamic values by following their appropriate FS-scaling (Fig.~\ref{fig:mag-mom}).

\subsection{DFT}
\label{DFT-methods}

While DFT calculations that are preliminary for our QMC simulations were done by using the Gaussian basis set code as implemented in the TurboRVB package (see Sec.~\ref{QMC-methods}),
all the other 
DFT\cite{hohenberg_inhomogeneous_1964} calculations were performed using the plane-wave (PW) based Quantum ESPRESSO\cite{giannozzi_quantum_2009,giannozzi_advanced_2017} package. 
The 2-ZGNR unit cell was defined as in the QMC calculations (see Fig.~\ref{fig:unit-cell}). At variance with QMC, no supercells are used in DFT. The vacuum separation in $y$ and $z$ directions was kept to 7.2 \AA~ and 7 \AA, respectively (see Fig.~\ref{fig:vac-z} of App.~\ref{DFT_appendix}).

One of the aims of this work is to study how DFT performs in ZGNRs as compared with benchmark QMC results. Various functionals have been used for this comparison:
LDA\cite{ceperley_ground_1980,perdew_self-interaction_1981}, LDA+U\cite{PhysRevB.52.R5467,PhysRevB.44.943}, the generalized gradient approximation in the Perdew–Burke–Ernzerhof implementation (GGA-PBE)\cite{perdew_generalized_1996}, GGA-PBE+U\cite{PhysRevB.57.1505}, the DFT-DF2 (van der Waals corrected) functional\cite{PhysRevB.82.081101}, the PBE functional revised for solids (PBEsol)\cite{PhysRevLett.100.136406}, the Becke-Lee-Yang-Parr (BLYP) functional\cite{PhysRevB.37.785}, the Gaussian-PBE (GauPBE)\cite{song_communication_2011}, Heyd–Scuseria–Ernzerhof (HSE)\cite{doi:10.1063/1.1564060,doi:10.1063/1.2204597}, and PBE0\cite{doi:10.1063/1.472933,adamo_1999_toward} hybrid fuctionals. 
In our PW-DFT calculations, the electron-ion interactions are described by the optimized norm-conserving Vanderbilt (ONCV) pseudopotentials\cite{PhysRevB.88.085117} for PBE and PBE-based hybrid functionals, while we used projector-augmented wave (PAW) pseudopotentials\cite{paw_pseudos} for LDA, LDA+U, PBE+U, PBEsol, and BLYP functionals.

A kinetic energy (charge density) cut-off of 80 Ry (320 Ry) was used in all PW calculations. This guarantees a sub-meV accuracy for both ONCV and PAW pseudopotentials. For the $\textbf{k}$-points integration, the Brillouin zone was sampled with 32 $\textbf{k}$-points in the ribbon direction, corresponding to a 
 $32 \times 1 \times 1$ Monkhorst-Pack grid\cite{monkhorst_special_1976}. We used a Marzari-Vanderbilt\cite{marzari_ensemble_1997} smearing of 0.006 Ry. This setup yields an accuracy below 1 meV in total energies (see App.~\ref{DFT_appendix} and Fig.~\ref{fig:smearing-convergence}).
 In the case of hybrid functionals, a downsampled $\textbf{q}$-grid of $16 \times 1 \times 1$ has been employed for the calculation of the exact-exchange operator, which again fulfills a target accuracy of 1 meV in total energies. 

The DFT+U calculations were performed within the 
Cococcioni's scheme to evaluate the Hubbard U operator \cite{cococcioni2005linear},
acting on the carbon $p$ states. This approach turns out to be equivalent to the Liechtenstein's formulation\cite{harmon_calculation_1995},
where we set the Hund's parameter $J=0$. Indeed, $\pi$-electron localization involves a single $p_z$ orbital per C site. In this configuration, the on-site intra-orbital Hubbard $U$ parameter is enough to fully characterize the Hubbard repulsion, and the two approaches coincide.

To check the computational accuracy of the aforementioned exchange-correlation functionals for 
the 2-ZGNR,
we performed different set of calculations. We computed magnetic and non-magnetic solutions of 2-ZGNR 
with:
\begin{itemize}
\item
fully relaxed DFT geometries within LDA, LDA+U, PBE, PBE+U, PBEsol and BLYP functionals. For the calculations with hybrid functionals, we used PBE-relaxed geometries;
\item
fully relaxed QMC geometries.
\end{itemize}
Geometry relaxation was converged with ionic forces below the $10^{-4}$ Ha/\AA\ threshold, pressure below $10^{-1}$ kbar.
The final geometries are reported in Tabs.~\ref{tab:table-pm-geo} and \ref{tab:table-afm-geo}. The energy stability of the AFM phase and its absolute magnetization are reported in Sec.~\ref{DFT-results}.

\section{RESULTS AND DISCUSSION}
\label{results}
\subsection{Ground-state properties from QMC}
\label{QMC-results}

 We performed QMC calculations for two different possible solutions of the 2-ZGNR ground state:  the unpolarized 
 PM phase and 
 the 
 AFM state. If for the GS of the system it is energetically more convenient to break the spin symmetry and become spin-polarized, the Lieb's theorem\cite{lieb1989two}, applied to a bipartite honeycomb lattice with on-site interactions, states that the solution must be antiferromagnetic. Thus, the PM energy needs first to be compared with the AFM one, in order to know whether the 2-ZGNR GS develops a long-range spin order, as predicted by previous DFT calculations. This question is particularly relevant for quasi-1D systems, where quantum fluctuations are strong and can destroy long-range charge and spin orders. Electron correlations on top of enhanced quantum fluctuations make the GS determination a hard task. We will see that electron correlations in 2-ZGNR are indeed not negligible, and this adds another layer of complexity to the problem.
 
 To tackle the study of 2-ZGNR, we used a fully-fledged quantum Monte Carlo approach, as described in Sec.~\ref{QMC-methods}. By first optimizing and then projecting the variational wave function in Eq.~\ref{wf}, we obtained the fixed-node GS energy, namely the best fixed-node DMC energy compatible with the nodes of the wave function before projection. We would like to highlight here that the starting LDA nodes, as defined by the LDA Kohn-Sham orbitals of the preparatory Gaussian DFT calculations, have been relaxed and optimized in the energy minimization step, in the presence of the correlated Jastrow factor. Thus, in the DMC step we used improved nodes which have a milder FN error. The combination of a full wave function optimization performed at the VMC level, followed by its further projection through the LRDMC 
 diffusion Monte Carlo algorithm, makes the QMC a very accurate framework in solid state physics\cite{Wagner_2016,saritas2017investigation,benali2020toward,nakano2020speeding,motta2020ground}.

During the wave function optimization step, not only we improved the electronic wave function parameters but also we relaxed the geometry of the system. Indeed, both wave function optimization and structural relaxation are based on the variational energy minimization. Thus, they can be performed simultaneously. By starting from the PBE geometry, we relaxed the symmetry-independent structural parameters till convergence. 
Our
QMC statistics 
is such that, at convergence, the bond lengths fluctuate around $\approx 0.02$\AA. The optimal geometries found at the VMC level for the PM and AFM phases are reported in Tabs.~\ref{tab:table-pm-geo} and \ref{tab:table-afm-geo}, respectively. It is worth noting that the variation between the two relaxed geometries mainly involves the C3-C6 vertical distance, and to a minor extent the C1-C6 and C-H bond lengths (see Fig.~\ref{fig:unit-cell}). The AFM phase has looser bonds linked to the outermost carbon sites of the zigzag chain. This affects the values of both C1-C6 and C6-H1 independent parameters. As we will see later, this is due to a quite strong electron localization, evidenced by the formation of 
sizable
local magnetic moments at the edges of the 2-ZGNR. As a consequence, the electron localization in the AFM state weakens the covalent chemical bonds around the terminal C sites, causing their elongation. Another concomitant effect is the widening of the $\widehat{\textrm{C2-C1-C6}}$ angle, leading to a larger C3-C6 distance in the AFM phase. This angle goes from 118.56(11) degrees in the PM state to 119.09(18) degrees in the AFM phase.

\begin{table}[t!]
\begin{ruledtabular}
\caption{\label{tab:table-pm-geo} Fully relaxed PM geometries (lengths in \AA) at different level of theory. We report only the values of independent parameters, from which the whole geometry can be fully reconstructed. By symmetry, C1-C6 $=$ C5-C6 $=$ C2-C3 $=$ C4-C3, C1-C2 $=$ C4-C5, and C6-H1 $=$ C3-H2.
}
\begin{tabular}{cccccc}
Bond length (\AA) & \multicolumn{1}{l}{QMC} & \multicolumn{1}{l}{PBE} & \multicolumn{1}{l}{BLYP} & \multicolumn{1}{l}{LDA+U} & \multicolumn{1}{l}{PBE+U} \\
\hline
\hline
C1-C6 & 1.391 & 1.402 & 1.408 & 1.353 & 1.373 \\
C1-C2 & 1.439 & 1.461 & 1.469 & 1.411 & 1.432 \\
C3-C6 & 2.769 & 2.811 & 2.821 & 2.701 & 2.745 \\
C6-H1 & 1.078 & 1.091 & 1.092 & 1.081 & 1.081  
\end{tabular}
\end{ruledtabular}
\end{table} 

\begin{table}[t!]
\begin{ruledtabular}
\caption{\label{tab:table-afm-geo} As in Tab.~\ref{tab:table-pm-geo} but for fully relaxed AFM geometries (lengths in \AA).}
\begin{tabular}{cccccc}
Bond length (\AA) & \multicolumn{1}{l}{QMC} & \multicolumn{1}{l}{PBE} & \multicolumn{1}{l}{BLYP} & \multicolumn{1}{l}{LDA+U} & \multicolumn{1}{l}{PBE+U} \\
\hline
\hline
C1-C6 & 1.398 & 1.404 & 1.408 & 1.362 & 1.381 \\
C1-C2 & 1.439 & 1.459 & 1.469 & 1.403 & 1.424 \\
C3-C6 & 2.799 & 2.811 & 2.822 & 2.703 & 2.746 \\
C6-H1 & 1.083 & 1.091 & 1.092 & 1.080 & 1.081 
\end{tabular}
\end{ruledtabular}
\end{table}

\begin{figure}[ht!]
\includegraphics[width=0.9\columnwidth]{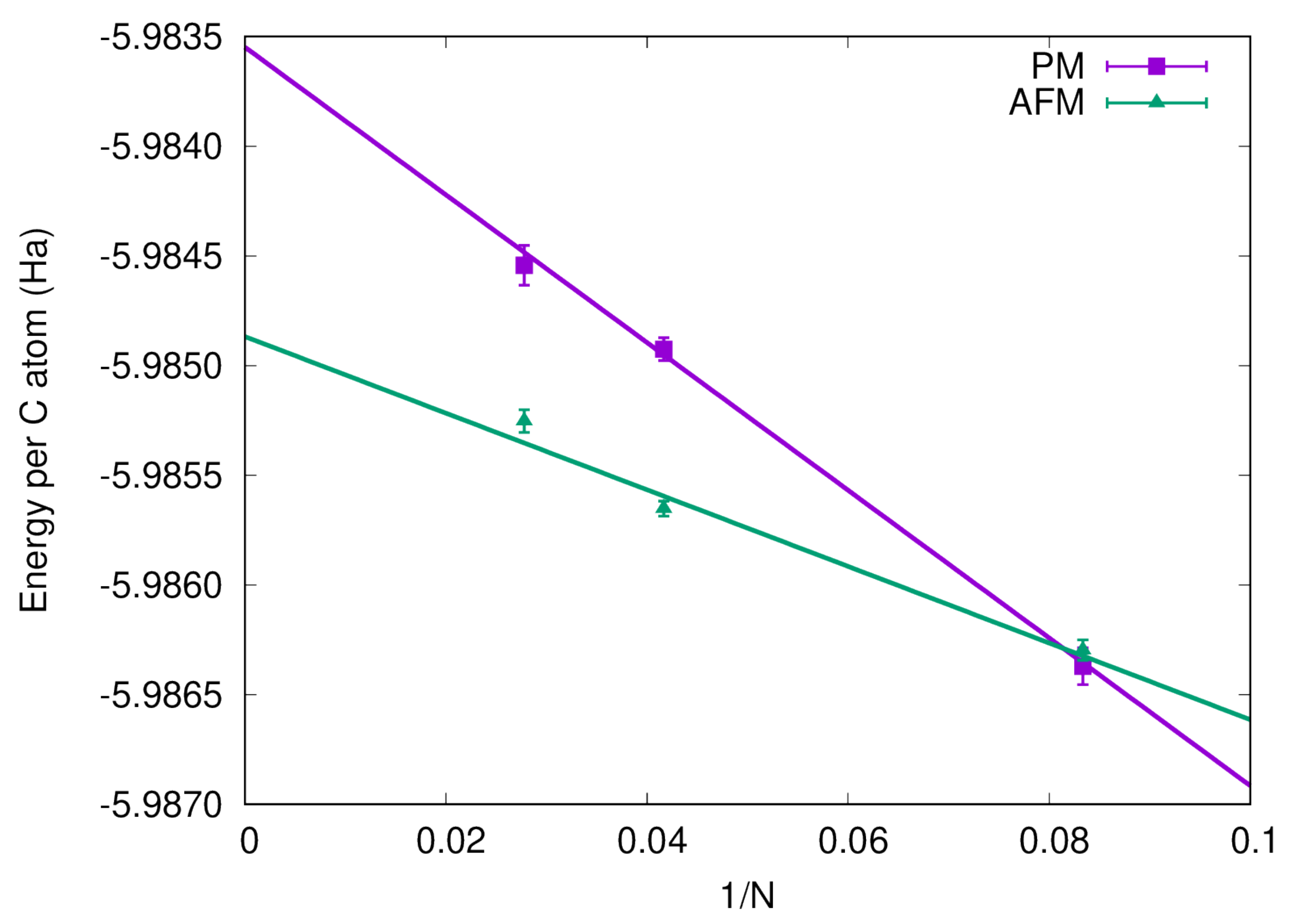}
\caption{\label{fig:FS-total-energy} Finite-size scaling of AFM (green) and PM (violet) fixed-node LRDMC energies per C atom.}
\label{fig:FN_energies}
\end{figure}

The FN-GS energies yielded by the PM and AFM wave functions computed at their respective relaxed geometries are plotted in Fig.~\ref{fig:FN_energies}, as a function of $1/N$. 
One can see that, 
although
the PM and AFM energies per C atom are nearly degenerate for a 3-ring supercell, the different slope of their finite-size scaling let the AFM energy be the lowest in the thermodynamic limit. Therefore, our QMC results confirm the early DFT finding of an antiferromagnetic ground state. Remarkably, the difference between the PM and AFM FN energies (AFM energy gain), once extrapolated to the thermodynamic limit, is large. It amounts to 36$\pm$3 meV per C atom (see Tab.~\ref{tab:afm-gain-qmc-geo}), significantly larger than previous predictions.

To give a more exhaustive explanation of this finding, let us analyze the two-dimensional (2D) spin magnetization density $\sigma_\textrm{2D}(x,y)=\langle \hat{\sigma}_\textrm{2D}(x,y) \rangle$, defined as the expectation value of the following operator:
\begin{equation}
\hat{\sigma}_\textrm{2D}(x,y)= - g \mu_B \sum_{i=1}^N   \hat{\sigma}_z(i) \delta(\hat{x}_i - x)\delta(\hat{y}_i -y),
\label{spin_density}
\end{equation}
where $\hat{\sigma}_z(i)$ is the $z$-component of the spin vector operator acting on the $i$-th particle spinor, $\hat{x}$ ($\hat{y}$) is the $x$ ($y$) component of the position operator, $\mu_B$ is the Bohr magneton, and the $g$-factor is taken equal to 2.  The LRDMC expectation value of the spin density operator in Eq.~\ref{spin_density} is computed over the mixed distribution, such that $\sigma^\textrm{LRDMC}_\textrm{2D}(x,y)=\langle \Psi_\textrm{FN} | \hat{\sigma}_\textrm{2D}(x,y) | \Psi \rangle/ \langle \Psi_\textrm{FN}| \Psi \rangle$. As usual for operators that do not commute with the Hamiltonian, this value does not coincide with the ``pure'' estimator of the FN ground state: $\sigma_\textrm{2D}(x,y)=\langle \Psi_\textrm{FN} | \hat{\sigma}_\textrm{2D}(x,y) | \Psi_\textrm{FN} \rangle$. We corrected for the mixed distribution bias by approximating the pure estimators through the linear-order correction in $O(|\Psi_{FN}-\Psi|)$, i.e.
$\sigma_\textrm{2D} \approx 2 \sigma^\textrm{LRDMC}_\textrm{2D} - \sigma^\textrm{VMC}_\textrm{2D}$, where $\sigma^\textrm{VMC}_\textrm{2D}(x,y)=\langle \Psi | \hat{\sigma}_\textrm{2D}(x,y) | \Psi \rangle$, the regular VMC expectation value.

\begin{figure}[ht]
\includegraphics[width=\columnwidth]{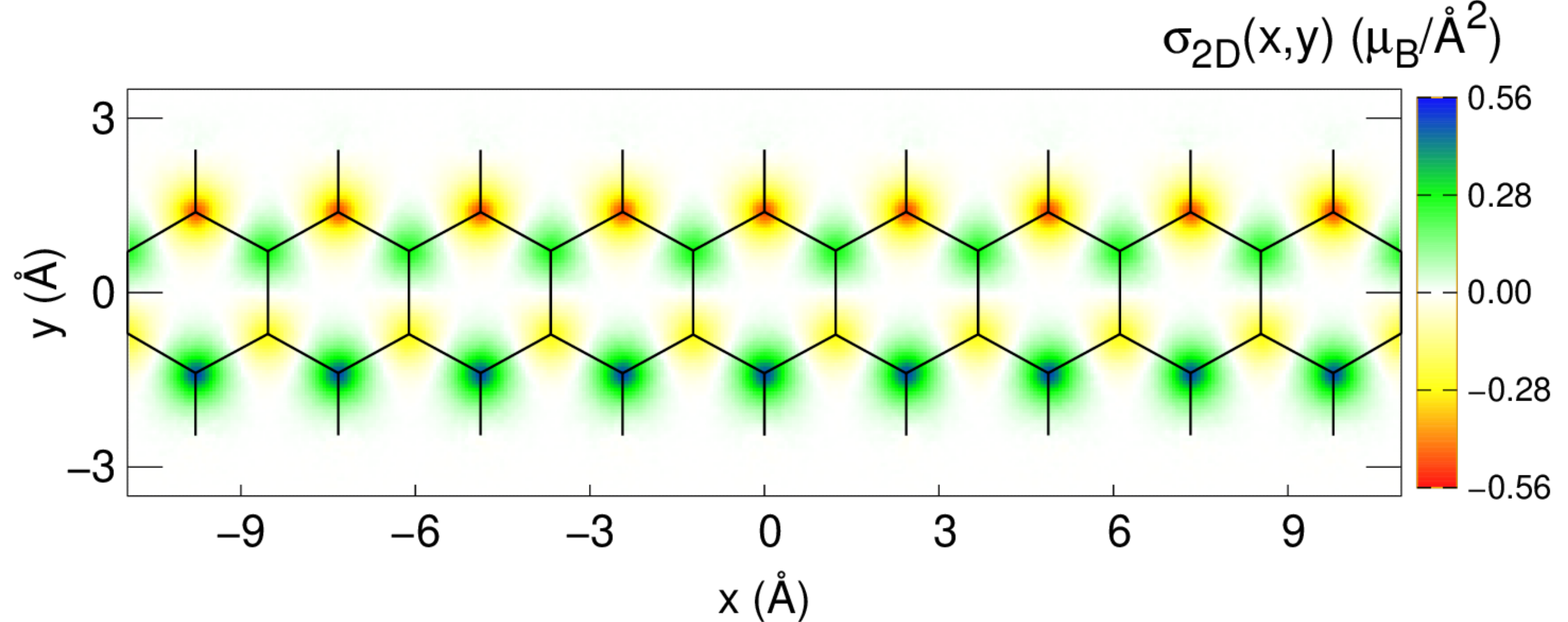}
\caption{\label{fig:contour-plot} Spin magnetization density 
for AFM fixed-node GS wave function in a 9-ring supercell. The quantum expectation values over the FN GS are corrected for the LRDMC mixed distribution (see text). Units are in $\mu_B$/\AA$^2$.}
\end{figure}

In Fig.~\ref{fig:contour-plot}, we plot $\sigma_\textrm{2D}(x,y)$ for the AFM fixed-node GS of a 9-ring supercell, which yields values close to the thermodynamic limit. It is apparent that the spins are arranged in an AFM pattern, as expected from the LSDA initialization of the AFM wave function (see Sec.~\ref{QMC-methods}). However, the outermost C sites host magnetic moments as large as 0.5 $\mu_B$, much larger than the initial LSDA values (see also Sec.~\ref{DFT-results}). The spread of the peaks in the $\sigma_\textrm{2D}(x,y)$ distribution show a high degree of localization of the magnetic moments. For the inner C sites, the spin moments are 
smaller than the terminal ones. The change of intensity between the outermost and the inner C atoms has also been reported in other ZGNR studies\cite{son_half-metallic_2006,PhysRevLett.99.186801,jiang_unique_2007,magda2014room}. Finally, in Fig.~\ref{fig:contour-plot} it can be seen that, in the AFM phase, the magnetic moments at the opposite edges are antiferromagnetically aligned, while across the ribbon direction the moments are ferromagnetically coupled. Instead, the PM phase does not show any pattern in $\sigma^\textrm{LRDMC}_\textrm{2D}(x,y)$, because the FN constraint 
hampers the spin-symmetry breaking during the projection,
and the corresponding spin density vanishes within the error bars.

\begin{figure}[ht]
\includegraphics[width=0.9\columnwidth]{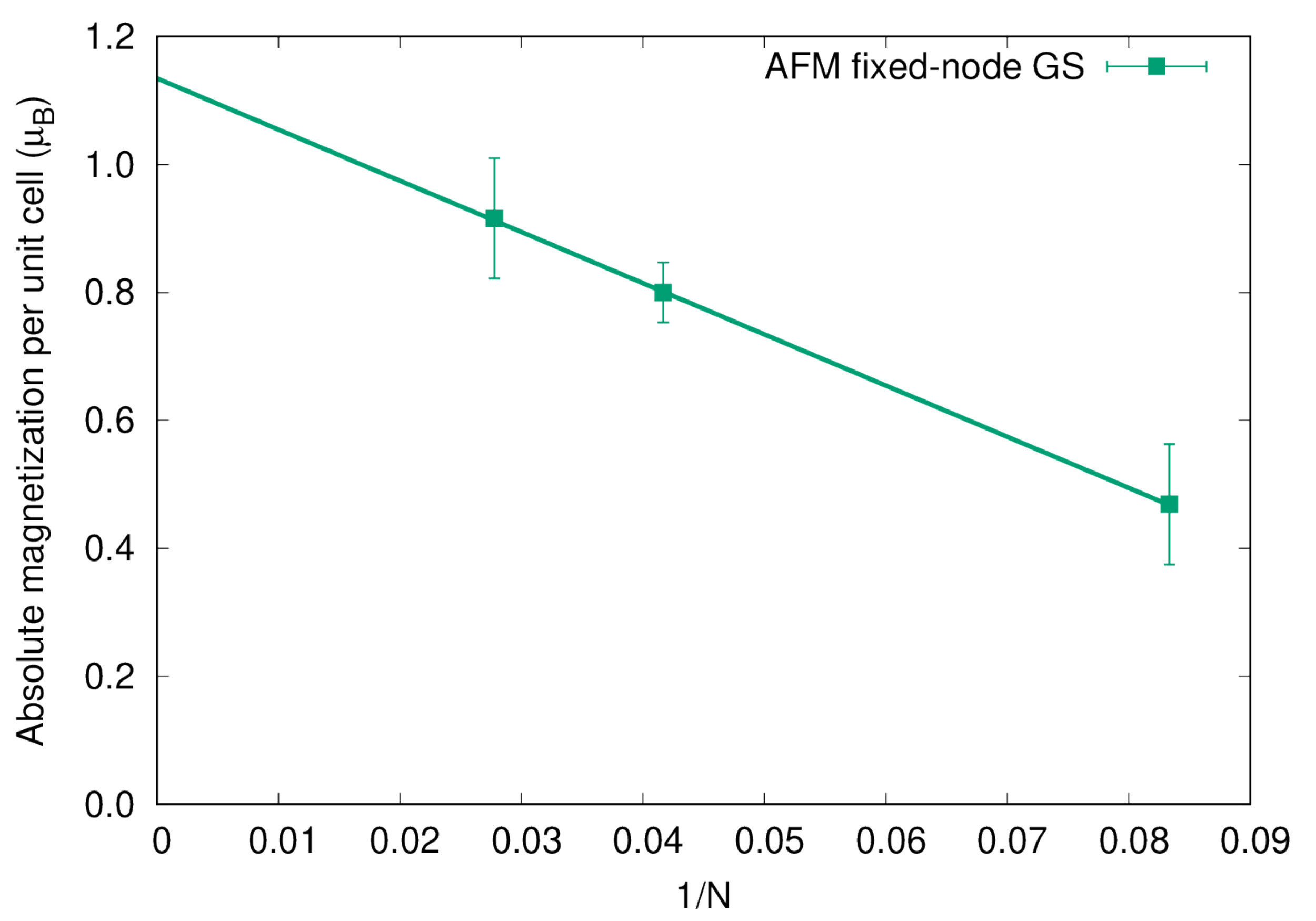}
\caption{\label{fig:mag-mom} Finite-size extrapolation of absolute magnetization per unit cell $M$, obtained for
the AFM FN ground state, corrected for the LRDMC mixed distribution.
}
\end{figure}

Fig.~\ref{fig:contour-plot} shows that the AFM pattern is a robust feature of the spin-broken LRDMC wave function. The intensity of this pattern is much more pronounced than in the LSDA calculations used to generate the initial wave function. Thus, electron correlations must enhance electron localization and local spin moment formation. To rigorously quantify this enhancement, we compute the absolute magnetization per unit cell $M$, defined as
\begin{equation}
    M = \int_V \!\textrm{d}x \textrm{d}y ~ |\sigma_\textrm{2D}(x,y)|,
\end{equation}
 where the integral is performed over the unit cell volume $V$. 
 $M/N_\textrm{C}$ gives the average magnetic moment per C atom in the cell. As we did for the spin magnetization density, we computed the value $M$ of the unbiased estimator for the LRDMC fixed-node wave function by means the first-order expression: $M=2M_\textrm{LRDMC}- M_\textrm{VMC}$. The finite-size scaling of $M$ is shown in Fig.~\ref{fig:mag-mom}. It linearly extrapolates to the value of 1.13 $\mu_\textrm{B}$, as also reported in Tab.~\ref{tab:afm-gain-qmc-geo}. 
 
 In Sec.~\ref{DFT-results}, we will compare our QMC results with the outcome of several DFT exchange-correlation functionals. We will see that the value of $M$ strongly correlates with the AFM energy gain. The larger $M$, the stronger the AFM pattern, and the more energetically stable the AFM GS is. This can easily be explained by assuming that the underlying model Hamiltonian governing the energetics and magnetism of the 2-ZGNR is the $t-J$ model restricted to the $p_z$ orbitals, where $t$ is the hopping, or hybridization strength, between two neighboring C sites, and $J$ is the spin-exchange coupling between nearest neighbors. Already at the mean-field level\cite{inui1988coexistence,jayaprakash1989mean,kyung2000new}, both energy and staggered magnetization of the $t-J$ model scale with $J$. Thus, they are correlated in the same way, and this correlation is stronger and stronger as the electrons get more and more localized. Indeed, strictly speaking, the $t-J$ representation holds in the strong electron localization limit. The positive linear relationship between the AFM energy gain and $M$  is also verified by their finite-size scaling behavior. Indeed, they both increase linearly as $N\rightarrow \infty$, as shown in  Figs.~\ref{fig:FN_energies} and \ref{fig:mag-mom}.

\subsection{Ground-state properties from DFT and comparison with QMC}
\label{DFT-results}

As explained in Sec.~\ref{DFT-methods},
we fully relaxed the geometries with LDA, LDA+U, PBE, PBE+U, PBEsol and BLYP functionals for both PM and AFM phases, whenever the latter exists as stable phase. For further calculations with hybrid functionals, we used the PBE relaxed geometries. The relaxed PM and AFM geometries are reported in Tabs.~\ref{tab:table-pm-geo} and \ref{tab:table-afm-geo}, respectively, for the most relevant functionals, while the corresponding PM-AFM energy difference (AFM energy gain) per C atom and AFM absolute magnetization per unit cell $M$ are reported in Tab.~\ref{tab:afm-gain-pbe-geo}, and compared to our QMC benchmark calculations. Moreover, for the sake of comparison, in Tab.~\ref{tab:afm-gain-qmc-geo} we report the AFM energy gain and $M$ values computed for different functionals at the same PM and AFM geometries taken from QMC and kept then frozen.

\begin{table}[ht]
\caption{\label{tab:afm-gain-pbe-geo} AFM energy gain per C atom with respect to fully relaxed PM phase, and AFM absolute magnetization per unit cell $M$. All the calculations were done at the DFT optimized geometries, except for reference QMC values. These data are plotted in Figs.~\ref{fig:afm-gain-mag-mom}(a) and \ref{fig:afm-gain-mag-mom}(b).
}
\begin{ruledtabular}
\begin{tabular}{lcc}
Level of theory & \multicolumn{1}{l}{AFM gain (meV)} & \multicolumn{1}{l}{ $M$ ($\mu_B$)} \\
\hline
\hline
BLYP         & 1.3  & 0.35 \\
PBE          & 3.4  & 0.53 \\
DFT-DF2      & 5.9  & 0.63 \\
HSE          & 24.2 & 0.92 \\
PBE0         & 34.0 & 1.02 \\
GauPBE       & 28.6 & 1.03 \\
LDA+U=7.6 eV & 39.1 & 1.12 \\
PBE+U=5.0 eV & 33.9 & 1.13 \\
\hline
QMC (reference)       & 36$\pm$3 & 1.13$\pm$0.01 
\end{tabular}
\end{ruledtabular}
\end{table}

\begin{figure*}
            \includegraphics[width=0.49\textwidth]{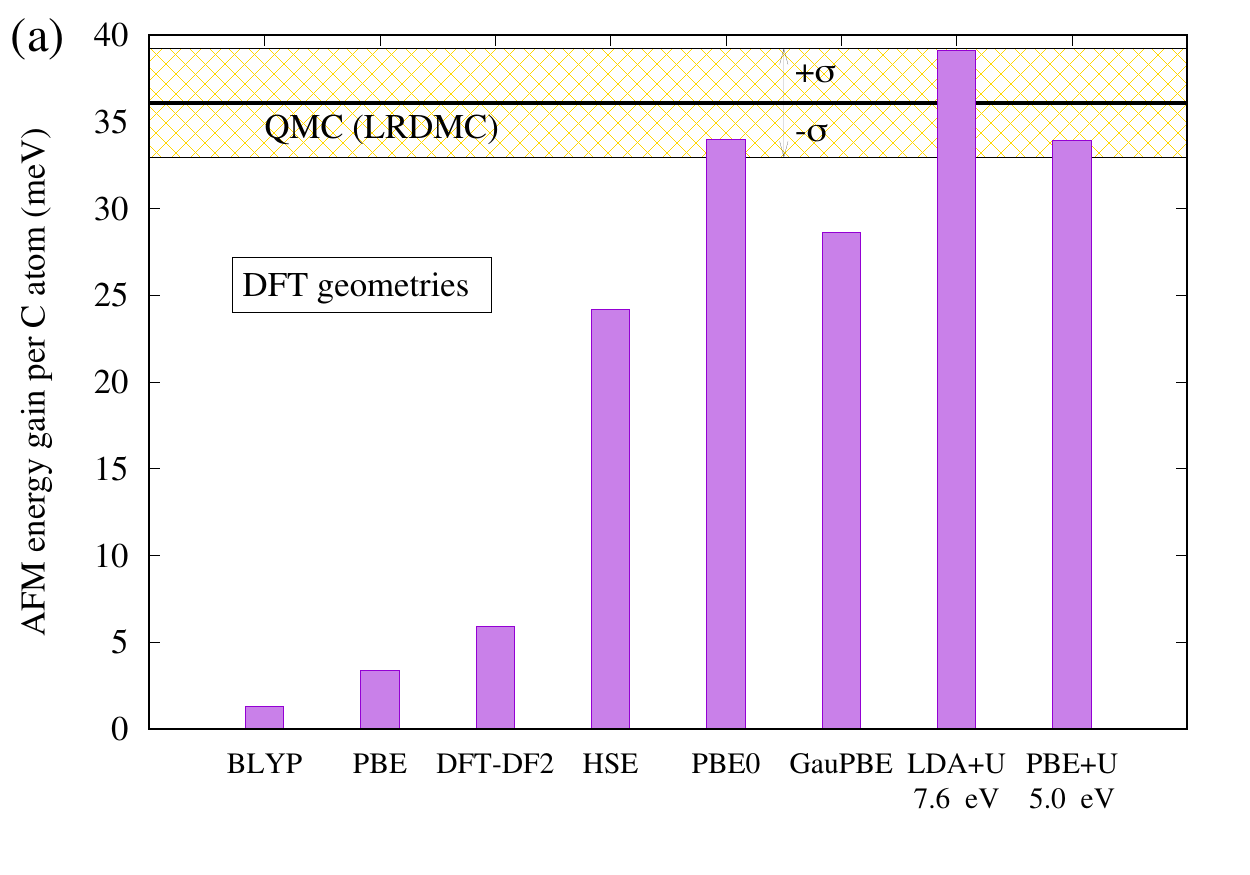}
            \includegraphics[width=0.49\textwidth]{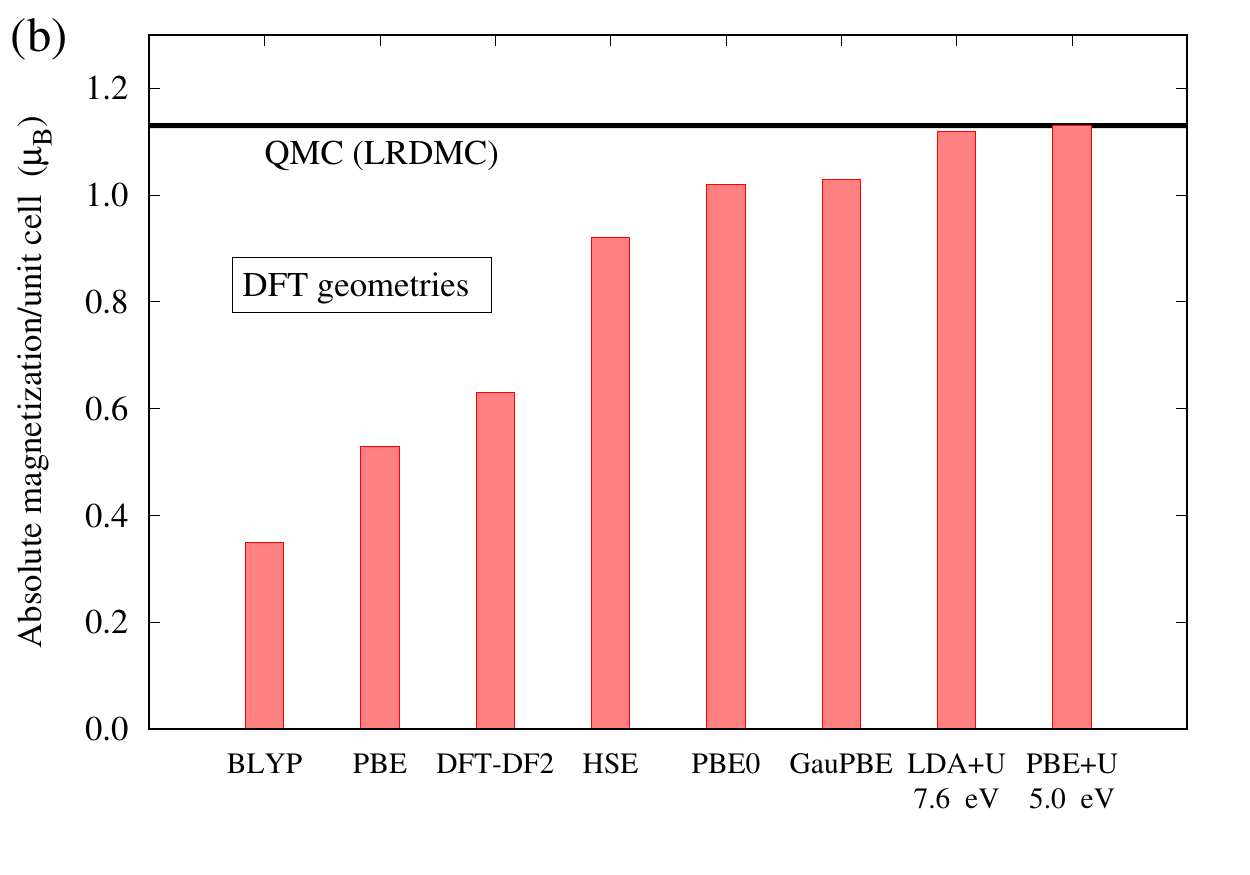}
            \includegraphics[width=0.49\textwidth]{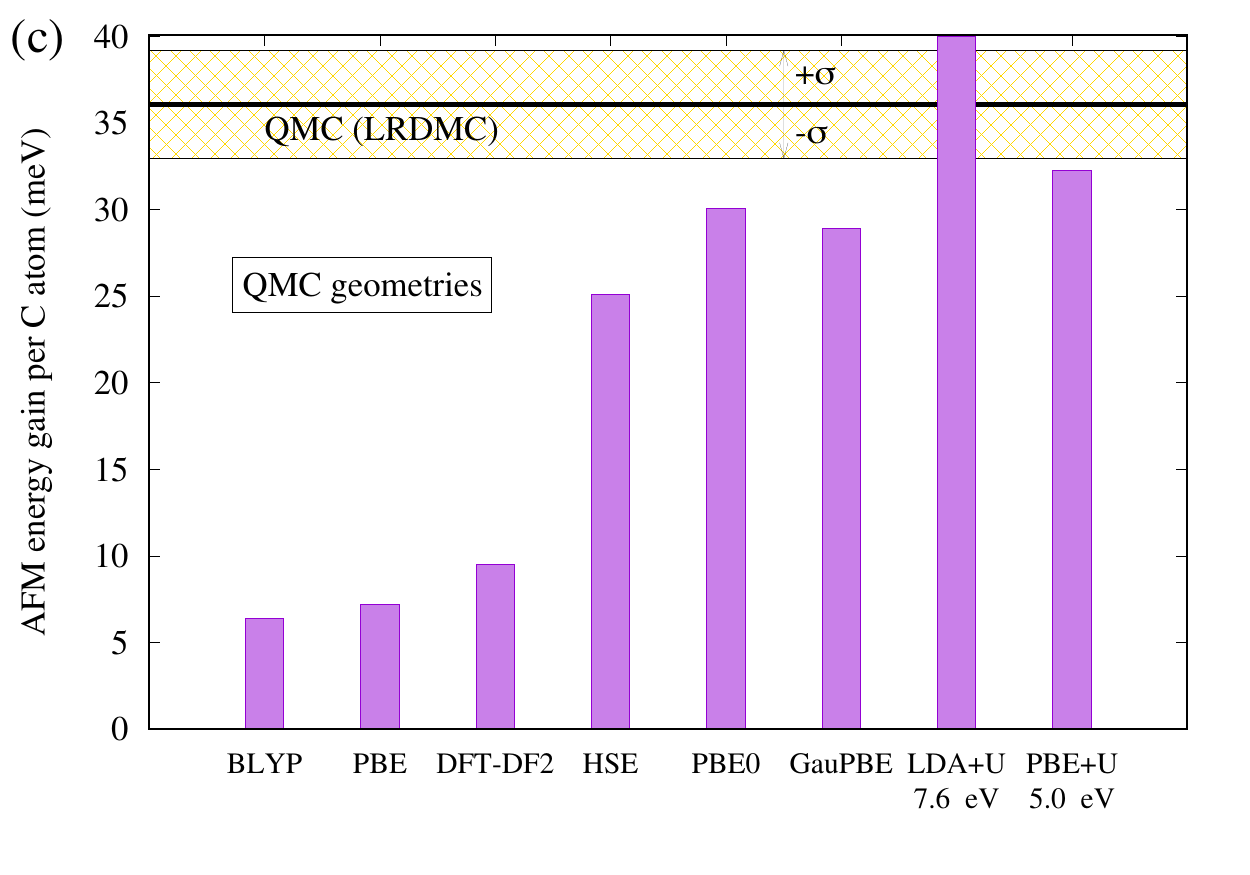}
            \includegraphics[width=0.49\textwidth]{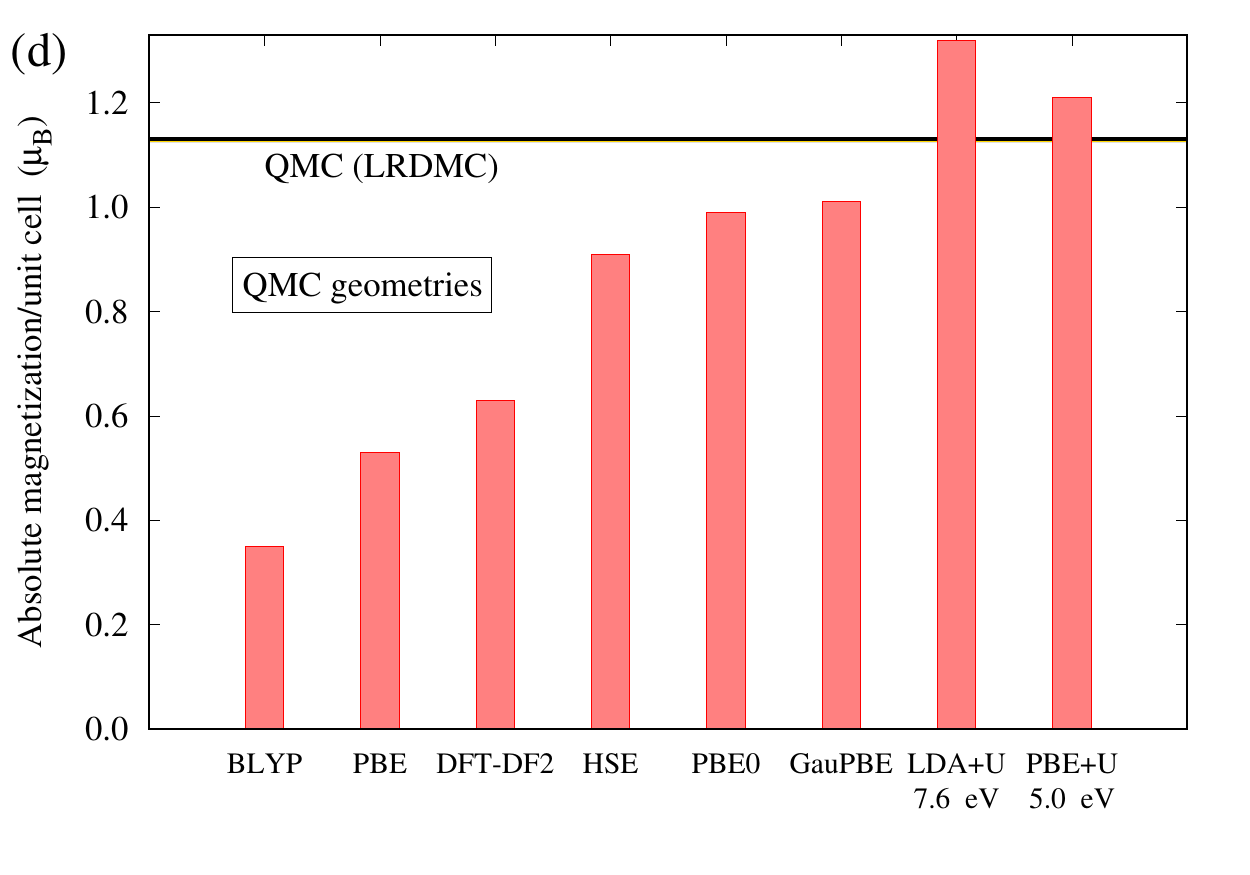}
\caption{\label{fig:afm-gain-mag-mom}
        (a) AFM energy gain defined as energy difference between PM and AFM phases, computed by different exchange-correlation functionals.
The value of the Hubbard repulsion U used in DFT+U calculations is reported at the bottom.
        The reference line corresponds to the value (36 meV) obtained at the QMC level. All the calculations are done at the corresponding DFT optimized geometries. (b) Absolute magnetization per unit cell 
        as a function of the same exchange-correlation functionals as used in panel (a).
        The reference line corresponds to the value (1.13 $\mu_B$) obtained at the QMC level. All the calculations are done at corresponding DFT optimized geometries. (c) 
The same as in (a) but for the QMC optimized geometries, taken same for all functionals.
        (d) 
The same as in (b) but for the QMC optimized geometries, taken the same for all functionals.
        } 
    \end{figure*}

Figs.~\ref{fig:afm-gain-mag-mom}(a) and \ref{fig:afm-gain-mag-mom}(b) plot the data in Tab.~\ref{tab:afm-gain-pbe-geo}. By analyzing them, it is apparent that in 2-ZGNR there is a strong correlation between the size of absolute magnetization and the AFM energy gain, as already pointed out. The larger the magnetization, the more stable the AFM phase is. In fact, 
in ZGNRs
$M$ is a proxy for the electron localization strength, as the local moment formation is directly related to the charge localization of a single $p_z$ electron per site, carrying a $1/2$ spin moment. The AFM exchange interaction $J$ between localized spins is then responsible for the lowering of the total energy in the AFM phase, and eventually for its gain with respect to the PM one. This is a strong coupling picture, where the Hubbard $U$ repulsion makes 
the particles more localized. Therefore, only \emph{ab initio} schemes able to deal with strong correlation are also capable of correctly describing the 2-ZGNR ground state.

It turns out that LDA and PBEsol fail to yield the AFM as stable phase, as they are, among the tested functionals, those that tend to delocalize the electrons the most. As a consequence, there is no stable AFM solution within these two functionals. It is worth noting that this result crucially depends on the correct $\textbf{k}$-point sampling of the IBZ. Indeed, if the $\textbf{k}$-mesh is not dense enough, i.e. it contains less than 30 $\textbf{k}$-points along the ribbon direction (see the $\textbf{k}$-point convergence plot of Fig.~\ref{fig:smearing-convergence}), one can still obtain a stable AFM solution in LDA. However, this results into a very fragile phase, i.e. with very small $M$ and low AFM energy gain\footnote{This is why we have been able to initialize an AFM wave function in the Gaussian LSDA framework for further QMC calculations.}. This could explain the previous LDA outcome published in Ref.~\onlinecite{correa_braiding_2018}, which reports an AFM stabilization energy of a few meV, while in our case the AFM gain is null. The sensitiveness to $\textbf{k}$-points sampling is clearly due to a Dirac cone formation in the PM band structure, arising from edge states\cite{correa_braiding_2018}. This is particulary evident in the 2-ZGNR, while the Dirac velocities flatten out for larger $n$-ZGNR, yielding a braided band structure.

At variance with LDA and PBEsol,
PBE and BLYP functionals
predict a stable AFM order, although the resulting phase has a too weak electron localization, and correspondingly a too low AFM energy gain, if compared to QMC.

Even the inclusion of dispersion interactions
through
the DFT-DF2 functional does not improve the AFM energy gain and absolute magnetization. This shows that dispersion interactions are not so significant in curing the correlation problem in ZGNR, as expected.

\begin{table}[ht]
\caption{\label{tab:afm-gain-qmc-geo} As in Tab.~\ref{tab:afm-gain-pbe-geo} but for QMC optimized geometries. These data are plotted in Figs.~\ref{fig:afm-gain-mag-mom}(c) and \ref{fig:afm-gain-mag-mom}(d).
}
\begin{ruledtabular}
\begin{tabular}{lcc}
Level of theory & \multicolumn{1}{l}{AFM gain (meV)} & \multicolumn{1}{l}{$M$ ($\mu_\textrm{B}$)} \\
\hline
\hline
BLYP         & 6.4  & 0.35 \\
PBE          & 7.2  & 0.53 \\
DFT-DF2      & 9.5  & 0.63 \\
HSE          & 25.1 & 0.91 \\
PBE0         & 30.1 & 0.99 \\
GauPBE       & 28.9 & 1.01 \\
LDA+U=7.6 eV & 40.0 & 1.32 \\
PBE+U=5.0 eV & 32.3 & 1.21 \\
\hline
QMC (reference)       & 36$\pm$3 & 1.13$\pm$0.01
\end{tabular}
\end{ruledtabular}
\end{table}

Since ``weakly correlated'' functionals, such as LDA, PBE, BLYP, PBEsol and DFT-DF2 functionals, fail to give an accurate description of the true GS, then use of 
hybrid functionals 
becomes
a 
natural
choice. 
Indeed,
hybrid functionals 
can treat electron correlation effects in a better way,
because of explicit incorporation of a portion of exact exchange from Hartree-Fock theory\cite{FISCHER1986274}. 
As a matter of fact,
the results obtained with HSE, PBE0 and GauPBE show a significant improvement over the LDA and GGA-based functionals.
PBE0 turns out to be the best hybrid functional for 2-ZGNRs, as it yields results close to QMC,
both in terms of AFM energy gain and $M$ values.
The reason of its apparent success
is the incorporation of fully non-local Hartree-Fock exchange\cite{adamo_1999_toward},
which favors the stabilization of 
broken-symmetry phases, such as the
AFM long-range order, as verified in ZGNR by model Hamiltonians\cite{jung2011nonlocal}. 
In HSE and GauPBE, the Hartree-Fock exchange is included only at short range, by using the error function and a Gaussian envelope as attenuation schemes\cite{song_communication_2011}, respectively, with the aim of screening the bare Coulomb interaction and making hybrid calculations more efficient.
They provide qualitatively similar results, with small variations arising from the different attenuation scheme employed. Nevertheless, on average, GauPBE performs slightly better than HSE in 2-ZGNR.

Despite their clear improvement with respect to weakly correlated functionals, hybrid functionals 
are not able to fully 
recover the QMC results. Therefore, in this work we explored 
another popular way to include strong correlation in DFT, i.e.
the explicit incorporation of the on-site Coulomb interaction U in the DFT+U approach.
In order to find the optimal value of U, we required that the AFM solution provided by DFT+U yields the \emph{same} $M$ value as QMC \emph{at relaxed geometries}. Indeed, electron localization probed by $M$ is a key quantity to assess the correlation level reached in the system. Requiring the same level of localization is physically more sound than choosing the AFM energy gain as target quantity, which could depend on how the density functional is built and defined. We noticed that geometry affects the $p_z$ electron localization. Thus, we relaxed the geometry within DFT+U at a given U value and we then computed the final corresponding $M$. In this way,
we 
found U=5.0 eV (U=7.6 eV) as optimal value in PBE+U (LDA+U).
Interestingly enough, these values 
are fully in range with those predicted by 
restricted random phase approximation (cRPA) calculations\cite{PhysRevB.98.205123} of ZGNRs, further strengthening our procedure. Indeed, site-specific cRPA estimates of the local U repulsion, based on the same PBE functional, show a reduction from the ``bulk'' value of graphene (U=9.3 eV)\cite{wehling2011strength,csacsiouglu2017strength} to the edge value of $\approx$ 5 eV, in nice agreement with our findings. In the narrowest 2-ZGNR, the edges play a dominant role in determining the site-averaged U repulsion, as the one we provide. It is also interesting to note that LDA+U requires larger U values than PBE+U. This agrees with a recent Bayesian calibration of Hubbard parameters in strongly correlated materials\cite{tavadze2021exploring}.

The DFT+U framework with optimal U values provides a further step forward with respect to the hybrid functionals for the 2-ZGNR description. LDA+U and PBE+U yield AFM energy gains that lie within one error bar from the best QMC estimates (Fig.~\ref{fig:afm-gain-mag-mom}(a)). At the same time, the $M$ values equal the QMC ones, thanks to the optimal-U construction (Fig.~\ref{fig:afm-gain-mag-mom}(b)).
The corresponding LDA+U and PBE+U geometries are reported in Tabs.~\ref{tab:table-pm-geo} and \ref{tab:table-afm-geo}. The agreement between DFT+U and QMC is less good for the geometries than for the energies. 

From Tabs.~\ref{tab:table-pm-geo} and \ref{tab:table-afm-geo}, it is clear that a highly correlated method such as QMC gives shorter equilibrium bond lengths than the PBE and BLYP functionals. However, DFT+U overestimates the bond length contraction. LDA+U severely overshoots the bond shortening, while this effect is much milder in PBE+U, which provides at the end much better geometries than LDA+U.

Another consequence of electron correlation 
is to enhance the structural differences between PM and AFM phases. While in PBE and BLYP the PM and AFM geometries are nearly the same, there is a significant variation in the QMC geometries across the phase change.
Upon inclusion of the on-site Coulomb interaction U, the difference between AFM and PM geometries becomes noticeable also in DFT+U. The C1-C6 and C3-C6 bond lengths in the AFM phase are longer than their counterparts in PM phase. This is true in both LDA+U and PBE+U, and it is in agreement with the QMC results.

Overall, supplementing the LDA and PBE functionals with local Hubbard U leads to equilibrium geometries that mimic more closely the structural behavior seen in QMC, with PBE+U outperforming LDA+U. 

To test the robustness of the results shown in Figs.~\ref{fig:afm-gain-mag-mom}(a) and \ref{fig:afm-gain-mag-mom}(b) against structural variations, we computed both AFM gain and absolute magnetization $M$ at the PM and AFM geometries borrowed from QMC and kept the same for all functionals. This makes the comparison with QMC more direct because it avoids a possible source of disagreement. The corresponding results are reported in Tab.~\ref{tab:afm-gain-qmc-geo} and plotted in Figs.~\ref{fig:afm-gain-mag-mom}(c) and \ref{fig:afm-gain-mag-mom}(d).

The qualitative picture does not depend on the actual geometries chosen. The local and semi-local functionals drastically fail, while hybrid and DFT+U functionals provide much more reliable results. Even quantitatively, the picture stays almost the same, with variations of a few meV between the AFM gains computed by using relaxed geometries and QMC geometries. The most relevant difference is the change of absolute magnetization $M$ in the DFT+U framework upon geometry variation. At the QMC geometry, both LDA+U and PBE+U yield larger $M$ than the QMC values.  Nevertheless, the PBE+U $M$ value is still very good, as it lies within 0.1 $\mu_B$ from the QMC reference. At the same time,
the PBE+U AFM gain is the closest to QMC among all tested functionals. The superior performances of PBE+U (U=5.0 eV) with respect to LDA+U (U=7.6 eV) and to the other functionals is thus verified for various independent properties: equilibrium geometries, AFM gain per C atom, AFM absolute magnetization per unit cell, robustness against geometry variation.

We finally looked for other possible instabilities arising from magnetism or geometry. Breaking the 
crystal
symmetry by introducing \emph{cis} and \emph{trans} type of distortions\cite{kivelson_polyacene_1983,dos_santos_electronic_2006} in the fused benzene rings at both PBE and 
GauPBE levels of theory does not lead to any structural instability in the AFM phase, for both functionals. Therefore, our results discard the existence of multiferroic ground states in neutral 2-ZGNR\cite{fernandez2008prediction,jung2009carrier}.
Instead, we did find structural instabilities in the PM phase. Nevertheless, the energy gained by introducing such distortions is very tiny ($< 0.1$ meV) as compared to the energy gain obtained by spin symmetry breaking. Thus, structural \emph{cis}/\emph{trans}- instabilities are irrelevant for the 2-ZGNR in its GS.

As far as the spin sector is concerned,
the ferromagnetic phase melts at the PBE level, and it is certainly not the GS 
in GauPBE,
being always bound from below by the AFM phase. This is in agreement with Lieb's theorem\cite{lieb1989two}.

The AFM symmetry breaking is by far the most robust among possible instabilities. Our findings, based on QMC and ``correlated'' functionals, support the picture of a 2-ZGNR GS with localized $\pi$ electrons and long-ranged AFM correlations.

\section{CONCLUSIONS}
\label{conclusions}

In this work, we studied the ground state properties of the 2-ZGNR by means of very accurate QMC calculations. We found that the best candidate for the 2-ZGNR ground state is a wave function developing an AFM long-range order at zero temperature. Despite the low dimensionality of the system, the AFM phase is very robust against spin quantum fluctuations, and the AFM magnetic pattern is stable in our LRDMC simulations for all ribbons with supercell size equal to or more than 3 fused benzene rings. This is at variance with the result obtained for the acene series\cite{dupuy_fate_2018}, the molecular analogues of the 2-ZGNR, where a paramagnetic wave function is the primary candidate for their ground state. The consequences of making the ribbon length \emph{finite}, by chopping a 2-ZGNR into an acene molecule, deserve further studies.

This work also provides QMC benchmark results which have been used to validate different DFT functionals, both in terms of AFM energy gain and AFM absolute magnetization. We 
tested
different DFT functionals such as LDA, LDA+U, PBE, PBE+U, DFT-DF2, BLYP, HSE, PBE0 and GauPBE functionals. For DFT+U frameworks, we determined the optimal value of U, thanks to the comparison with our QMC results.
We can conclude that the PBE+U functional with U=5.0 eV is the best among all the DFT functionals reported in this work, in terms of geometry, electron localization, magnetic moment, and AFM stabilization energy. The optimal U repulsion strength in PBE+U is in a very good agreement with its cRPA-PBE determination\cite{PhysRevB.98.205123}, and with a previous estimate based on the measured magnitude of ZGNR gaps and on the semiconductor–metal transition ZGNR width, found experimentally\cite{magda2014room}.
Hence, it would be rather safe to extend the use of PBE+U 
from 2-ZGNR to computing materials properties of analogous C-based systems, with a graphene pattern: nanotubes, nanoribbons, 
graphene impurities with doping and magnetism. 

To conclude,
both
QMC and ``correlated'' functionals support the picture of a significant stabilization energy
of the AFM long-range order at zero temperature. This energy gain is much stronger than what predicted by previous DFT calculations using local or semi-local functionals. The quantitative failure of ``weakly correlated'' schemes is spectacular here. The AFM stabilization energy predicted by LRDMC, hybrids functionals and DFT+U schemes is from 5 to 7 times larger than the PBE one, and the associated local moments are from 2 to 3 times larger. 
GW calculations suggested that many-body correlation effects could make the ZGNR band gaps 2-3 times larger than in PBE\cite{yang2007quasiparticle}. In this work we show how a non-perturbative many-body approach such as QMC further enhances the tendency to AFM ordering in 2-ZGNR, entirely from first principles. Previous hints based on non-perturbative many-body calculations came mainly from the solution of the Hubbard model on a honeycomb lattice, where the choice of the effective on-site repulsion remains critical for quantitative estimates\cite{wehling2011strength,schuler2013optimal}. 
The remarkably strong stability of the AFM phase found in the 2-ZGNRs at zero temperature by \emph{ab initio} QMC techniques points towards the possibility of having stable antiferromagnetism above room temperature in this class of $\pi$-conjugated materials.

\begin{acknowledgments}
We are indebted to Prof. Prasenjit Ghosh, who took part in the early stage of the project, when RM was enrolled in the BS-MS Dual Degree Program involving the Department of Physics, Indian Institute of Science Education and Research (IISER), Pune, India, and the Institut de Min\'eralogie, de Physique des Mat\'eriaux et de Cosmochimie (IMPMC), Sorbonne Université, Paris, France. Prof. Prasenjit Ghosh acted as IISER supervisor.
We thank the European union for providing the Erasmus+ International Credit Mobility Grant for carrying out this collaborative project. RM would like to thank IISER Pune for providing a local cluster, Prithvi, which was used for preliminary DFT calculations. RM would also like to thank Wageningen University and Dutch National Supercomputing agency SURFsara for providing access to the Cartesius machine under the project EINF-750.
MC thanks GENCI for providing computational resources under the grant number 0906493, the Grands Challenge DARI for allowing calculations on the Joliot-Curie Rome HPC cluster under the project number gch0420, and RIKEN for the access to the Hokusai Greatwave supercomputer with the account number G19030. 
This work was partially supported by the European Centre of Excellence in Exascale Computing TREX—Targeting Real Chemical Accuracy at the Exascale. This project has received funding from the European Union’s Horizon 2020 Research and Innovation program under Grant Agreement No. 952165.
\end{acknowledgments}

\appendix
\label{appendix}

\section{\textbf{k}-points, smearing and box-size convergence in DFT}
\label{DFT_appendix}

\begin{figure}[ht]
        \centering
        \includegraphics[width=0.49\textwidth]{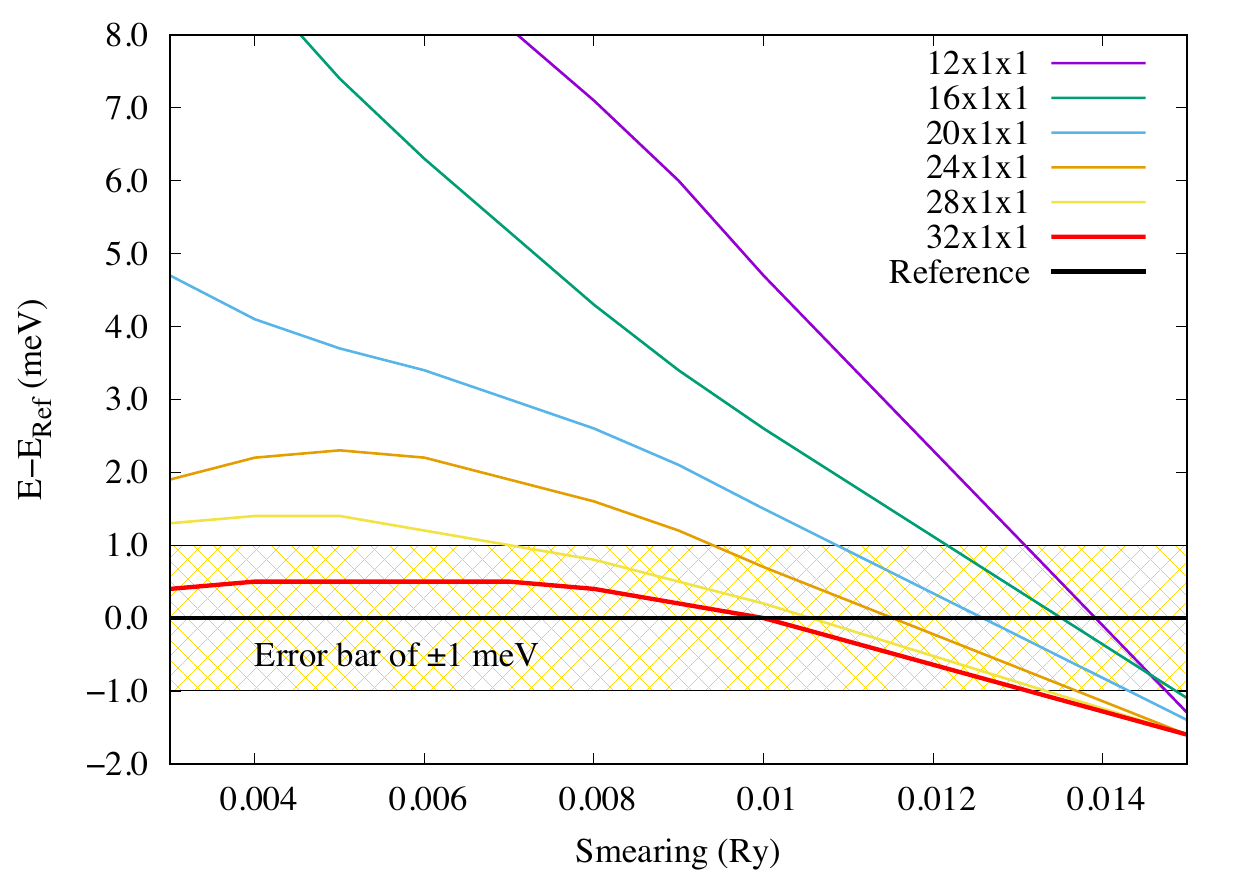}
        \caption{Total-energy (E) convergence as a function of smearing and $\textbf{k}$-mesh. The solid black line corresponds to the Reference energy (E$_\textrm{Ref}$) calculated with vanishing smearing
        and a $50 \times 1 \times 1$ $\textbf{k}$-mesh. The other colored lines connect points computed with
        the Marzari-Vanderbilt\cite{marzari_ensemble_1997} smearing, whose value is reported in the $x$-axis. The $y$-axis origin is pinned at E$_\textrm{Ref}$. The shaded yellow area is the target accuracy range.}
        \label{fig:smearing-convergence}
    \end{figure}

We study the total energy convergence as a function of smearing and number of $\textbf{k}$-points, in PBE calculations. By taking 1 meV as target accuracy, Fig.~\ref{fig:smearing-convergence} shows that the convergence condition is met for a $32 \times 1 \times 1$ Monkhorst-Pack $\textbf{k}$-grid, with the corresponding energy curve which levels off at a Marzari-Vanderbilt\cite{marzari_ensemble_1997} smearing of 0.006 Ry.
    
\begin{figure}[ht]
        \centering
        \includegraphics[scale=0.7]{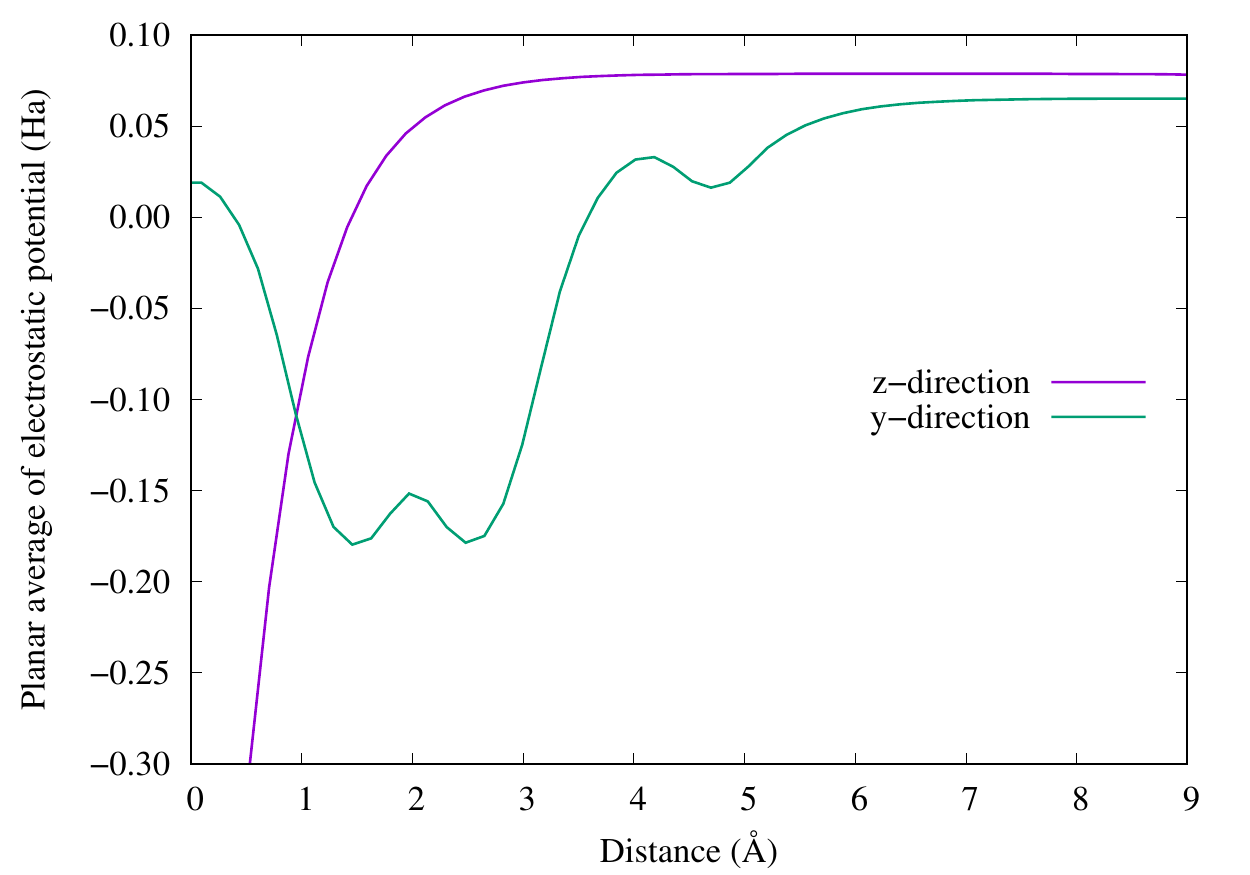}
        \caption{Planar average of electrostatic potential as a function of the distance in $y$- and $z$-direction from the axis of the 2-ZGNR, which lies in the $xy$ plane and is oriented in the $x$-direction (see Fig.~\ref{fig:unit-cell}).}
        \label{fig:vac-z}
    \end{figure}

The minimal vacuum distance that does not introduce spurious interactions between replicas in the PW-DFT framework has been estimated through the calculation of the planar average\cite{Fall_1999} of the electrostatic potential (V$_\textrm{bare}$+V$_\textrm{H}$) in $y$- and $z$-directions. Supercell self-consistent calculations provide the electronic charge density and the corresponding electrostatic potential. 
An acceptable 
vacuum distance along a certain direction is given by the position where the planar average of the electrostatic potential in that direction is flat.
A distance of 7 {\AA} (7.2 {\AA}) along the $y$ ($z$) direction from the center of the system fulfills this condition, as reported in Fig.\ref{fig:vac-z}.

\section{QMC total energies for different phases and sizes.}
\label{QMC_appendix}

We report the VMC and FN-LRDMC energies per C atom as a function of the supercell size, expressed as the number of C atoms in the supercell.  In the QMC Hamiltonian, the C atom is replaced by the corresponding pseudoatom as defined by the BFD pseudopotentials\cite{burkatzki_energy-consistent_2007}. We extended the system in the $x$ direction, by including 3, 6, and 9 fused benzene rings in the simulation box. Values for both PM and AFM wave functions are reported in Tabs.~\ref{tab:pm-QMCenergies} and \ref{tab:afm-QMCenergies}, respectively. These energies are KZK-corrected\cite{kwee_finite-size_2008}, and they correspond to the special $\textbf{k}$-point\cite{dagrada_exact_2016}, shown in the Tables.

\begin{table}[ht]
\begin{ruledtabular}
\caption{\label{tab:pm-QMCenergies} KZK-corrected QMC energies per C atom for the PM phase. $\textbf{k}_s$ is expressed in crystalline units, where $a$ is the length of the supercell in the $x$ direction.}
\begin{tabular}{cccc}
\multicolumn{1}{l}{\#C atoms} &
\multicolumn{1}{l}{$\textbf{k}_s$ ($2\pi/a$)} &
\multicolumn{1}{l}{$E_\textrm{VMC} (\sigma_\textrm{VMC})$(Ha)} 
& \multicolumn{1}{l}{$E_\textrm{FN} (\sigma_\textrm{FN})$(Ha)} 
\\
\hline
\hline
12  & 0.2435 & -5.975109(30)  & -5.986371(84) \\
24  & 0.1721 & -5.972818(9)   &  -5.984925(52) \\
36  & 0.2369 & -5.972040(12)  &  -5.984543(91) 

\end{tabular}
\end{ruledtabular}
\end{table}

\begin{table}[ht]
\begin{ruledtabular}
\caption{\label{tab:afm-QMCenergies} 
As in Tab.~\ref{tab:pm-QMCenergies}, but for the AFM phase.
}
\begin{tabular}{cccc}
\multicolumn{1}{l}{\#C atoms} &
\multicolumn{1}{l}{$\textbf{k}_s$ ($2\pi/a$)} &
\multicolumn{1}{l}{$E_\textrm{VMC} (\sigma_\textrm{VMC})$(Ha)} 
& \multicolumn{1}{l}{$E_\textrm{FN} (\sigma_\textrm{FN})$(Ha)} 
\\
\hline
\hline
12  & 0.2435 & -5.976840(14)   &  -5.986297(46) \\
24  & 0.1721 & -5.976176(9)    &  -5.985652(34) \\
36  & 0.2369 & -5.975603(10)   &  -5.985253(52) 
\end{tabular}
\end{ruledtabular}
\end{table}

\bibliography{JCP}

\end{document}